\documentclass[aps, preprint, showpacs]{revtex4}
\usepackage[T1]{fontenc}
\usepackage[latin9]{inputenc}
\usepackage{color}
\usepackage{amsmath}
\usepackage{graphicx}
\usepackage{esint}
\usepackage{ulem}



\begin{document}

\title{Hub-activated signal transmission in complex networks}

\author{Sven Jahnke$^{1-3}$, Raoul-Martin Memmesheimer$^{4}$, and Marc
Timme$^{1-3}$}

\affiliation{$^{1}$Network Dynamics, Max Planck Institute for Dynamics \& Self-Organization
(MPIDS), 37077 Göttingen, Germany,}

\affiliation{$^{2}$Bernstein Center for Computational Neuroscience (BCCN), 37077
Göttingen, Germany,}

\affiliation{$^{3}$Institute for Nonlinear Dynamics, Fakultät für Physik, Georg-August-Universität
Göttingen,}

\affiliation{$^{4}$Donders Institute, Department for Neuroinformatics, Radboud
University, Nijmegen, Netherlands.}

\date{\today}
\begin{abstract}
A wide range of networked systems exhibit highly connected nodes (hubs)
as prominent structural elements. The functional roles of hubs in
the collective nonlinear dynamics of many such networks, however, are not well
understood. Here we propose that hubs in neural circuits may activate
local signal transmission along sequences of specific subnetworks.
Intriguingly, in contrast to previous suggestions of the functional roles
of hubs, here not the hubs themselves, but non-hub subnetworks transfer
the signals.  The core mechanism relies on hubs and non-hubs providing 
activating feedback to each other. It may thus 
induce the propagation of specific pulse and rate signals in 
neuronal and other communication networks. 
\end{abstract}


\maketitle

Hubs -- nodes that are significantly more highly connected than average
-- constitute a prominent structural feature of many network dynamical
systems such as infection, transportation, communication and social
networks \cite{Newman}. The existence of hubs may
follow from intentional design to optimize network properties (such
as in airline, transportation and technical communication infrastructure)
or may emerge due to self-organization via intrinsic growth rules (World
Wide Web and social networks) \cite{Bornholdt,Newman,Kaluza2010,Barthelemy2011}.
As hubs can structurally improve the capabilities of networks to transfer
signals \cite{HubsSpreader}
it is not surprising that they were also found in the brain on different
scales: In cortical neuronal circuits, hub-regions are assumed to
coordinate the activity of other regions and organize the flow of
information between them \cite{HubRegions}.
On the microscopic level, for instance, the nervous system of  \textit{C.~elegans}
contains single cell hubs \cite{CElegans}
involved in the control of pheromone attraction as well as social
behavior \cite{Macosko2009}. Interestingly, Bonifazi et al. \cite{Bonifazi2009}
recently experimentally discovered hub cells also in higher animals
where they support synchronous activity in developing hippocampus.
Yet, how exactly hubs dynamically influence information transmission
in neural circuits still remains unknown \cite{Luccioli2013}.

In this Letter, we show that hub activity may amplify local signals
and enable their targeted transmission. Specifically, we show how
hubs and non-hub subnetworks in neural circuits activate
each other to exhibit synchronous pulse emission. Thereby, synchronous
pulse activity may robustly propagate along sequences of non-hub subnetworks,
thus enabling directed and specific routing of information across
the entire system. The generic mechanism of mutual hub and non-hub
activation may equally enable the transmission of pulse-coded as well
as rate-coded signals in a wide range of natural and artificial communication
networks. 

For an example of spiking neural circuits consider networks of $N$
units randomly connected to each other. 
Each connection is present with a fixed probability probability. 
In the simplest setting, between any pair of neurons there is 
an excitatory connection  of strength $\epsilon_{+}$ with 
probability $p_{+}$ and additionally an inhibitory 
connection of strength $\epsilon_{-}$ with probability $p_{-}=p_{+}=:p$.   
The dynamics of each unit $i$ is described by a real state variable,
its membrane potential $V_{i}(t)$, in real time $t$ and changes
according to leaky integrate-and-fire dynamics. Specifically,
$V_{i}$ integrates excitatory (positive) and inhibitory (negative)
pulsed inputs and when crossing a threshold from below, the potential
resets and the unit emits a pulse. 
This pulse arrives at the postsynaptic neurons after a
transmission delay and its effects are modeled by transient double-exponential 
conductance changes \cite{SI}.

Typically some of the pulse inputs to a neuron are synchronous (i.e.,
are received within a few milliseconds) and others are asynchronous.
Whereas the neuron integrates all inhibitory and asynchronous excitatory
inputs additively, synchronous excitatory inputs are processed non-additively
(non-linearly). This non-additive integration takes into account the
influence of fast dendritic spikes found in single neuron experiments
\cite{DSExp} on the
dendritic (input) sites of neurons: Whenever the total excitatory
input to a dendrite summed over a short time interval (typically 2-3~ms)
exceeds a dendritic threshold $\Theta_{\text{d}}$, a dendritic spike
is initiated and changes the membrane potential of the neuron after
a short delay in a stereotypical way. We model its effect by a stereotypical
current pulse causing a rapid, strong increase (depolarization) of
$V_{i}$, which substantially exceeds the level of depolarization
expected from linear summation of inputs \cite{Memmesheimer2010a,DSSynchProp,SI} and resembles the shape of the depolarization found in experiments \cite{DSExp}. 
We account for the experimentally observed saturation of the depolarization 
by inputs exceeding the dendritic threshold $\Theta_\text{d}$ \cite{DSExp}
as well as for the refractory time of ion channels generating
 dendritic spikes by assuming that the dendrite becomes refractory for a short time 
period $t^{\text{ref,ds}}$ after a dendritic spike is initiated.

In our numerical simulations, we focus on networks of spiking leaky integrate-and-fire neurons as described above. To achieve a mechanistic understanding of
the observed phenomena, we further derive an analytically tractable description in terms of probabilistic threshold units
below. 

Motivated by recent anatomical and physiological findings
  \cite{Bonifazi2009}, we assume that some $N_{\text{h}}\geq0$
neurons are hub neurons. They 
are distinguished (exclusively) by an increased probability $p_{\text{h}}>p$
to receive input connections from other units in the network.

Following a standard approach for signal transmission in cortical networks \cite{Synfire}, 
we consider signal propagation along weak feed-forward structures: 
The network contains sequences (chains) of $m$ subnetworks (groups)
with $N_{\text{g}}$ neurons each. 
The neurons in each group are randomly chosen 
from the non-hub population and excitatory connection
strengths between subsequent subnetworks are
  increased compared to other coupling
strengths in the network, $\epsilon_{\text{sub}}>\epsilon_{+}$. 

\begin{figure}[!tb]
\centering{}\includegraphics[width=0.6\columnwidth]{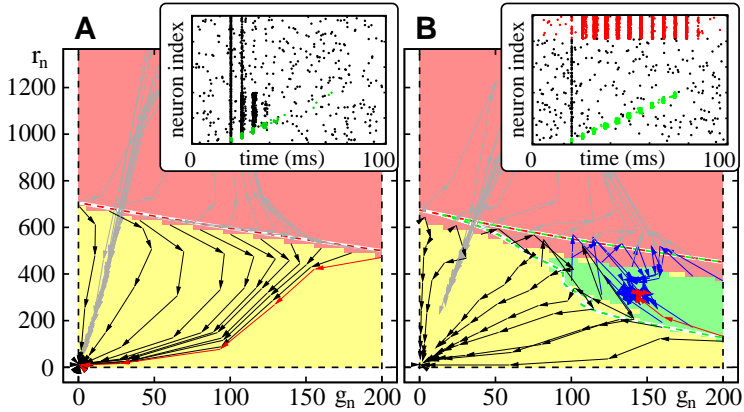}\caption{Hubs
  activate signal transmission in a neural network.
Signals consist of localized synchronous spiking activity (times marked
green in insets) transmitted across a sequence of subnetworks (displayed
as lowest neuron indices). Spike times of hubs (red) displayed at the top, 
above those of the remaining neurons (black). Main panels: 
Joint dynamics of the number of synchronously spiking neurons in 
the $n$th subnetwork ($g_n$) and 
the total number of synchronously co-activated neurons of the network 
remainder ($r_n$) during signal propagation initiated by synchronously 
stimulating $g_0$ neurons of the initial subnetwork and $r_0$ 
neurons of the network remainder. (A) In networks
without hubs, the overall network activity either becomes pathological
(large scale synchrony: red shading, gray trajectories) or extinguishes
quickly to background activity (yellow shading, black trajectories).
Hub-neurons in otherwise the same network (B) can  induce a persistent
signal transmission across non-hubs (green shading, blue trajectories)
by generating sustained but bounded synchrony. Red trajectories indicate
example  dynamics shown in insets. Dashed lines indicate the borders
of activity regions analytically estimated in this article (cf.~Eqs.~(\ref{eq:gnp1},\ref{eq:hnp1})
and \cite{SI}). 
Parameters: $N=5000$, $m=10$, $N_\text{g}=200$, $p=0.05$; further $N_\text{h}=0$ in (A) and $N_\text{h}=900$, $p_\text{h}=0.12$ in (B).
\label{fig:Hubs-enable-propagation}}
\end{figure}

We consider networks with balanced excitatory and inhibitory connectivity, such that
in the absence of external inputs, asynchronous irregular spiking
dynamics constitutes their ground state activity \cite{Balanced}.
Externally exciting an initial subnetwork to spike synchronously
causes synchronous inputs to neurons of the downstream subnetwork
and induces  synchronous spiking of a fraction of its neurons.
This may excite neurons in
the ensuing subnetwork to spike etc., thereby
transmitting signals along the chain of subnetworks. However, as
the subnetworks are parts of a larger recurrent network, synchronous
activity may spread not only from one subnetwork to the next, but
also induce a synchronous spiking response (echo) in the remainder
of the network. Depending on parameters and the number of initially
synchronous neurons $g_{0}$ in the first subnetwork and $r_{0}$ in the
remainder of the network, synchronous activity may in principle stably propagate, spread across
the entire recurrent network and thus obscure a propagation signal
(not shown) or extinguish after a few subnetworks.

Sample simulations
of networks without hubs ($N_{\text{h}}=0$, Fig.~\ref{fig:Hubs-enable-propagation}A)
illustrate that spreading and dying out of synchrony dominate state
space, in agreement with the literature \cite{Embed},
because there is no mechanism keeping the synchronization in the network remainder at a moderate level.

Networks with a substantial number $N_{\text{h}}$ of hub units exhibit
qualitatively different dynamics and support signal transmission:
As hubs receive more input connections than other units they have
a higher probability of spiking in response to synchronous inputs
from a certain subnetwork. Thereby, hubs may establish a synchronous
response to propagating synchronous pulses. Due to increased connectivity
at hubs only, such an echo is confined to the hub neuron sub-population
and thus does not spread over the entire network (cf.~Fig.~\ref{fig:Hubs-enable-propagation}B).

The increased connectivity towards hubs plays an interesting double
role: It ensures that a population of sufficiently many hub neurons
exhibits itself synchronous activity if supported by synchrony in a (non-hub) subnetwork. 
At the same time, the fact that the network remainder
without hubs has relatively low connectivity prevents spreading of
synchronous activity beyond the hub population. This combination enables
robust synchrony propagation along sequences of non-hub subnetworks
for a range of initially synchronous neurons $g_{0}$ in a subnetwork
(cf.~Fig.~\ref{fig:Hubs-enable-propagation}B, main panel).

\begin{figure}[!tb]
\centering{}\includegraphics[width=0.6\columnwidth]{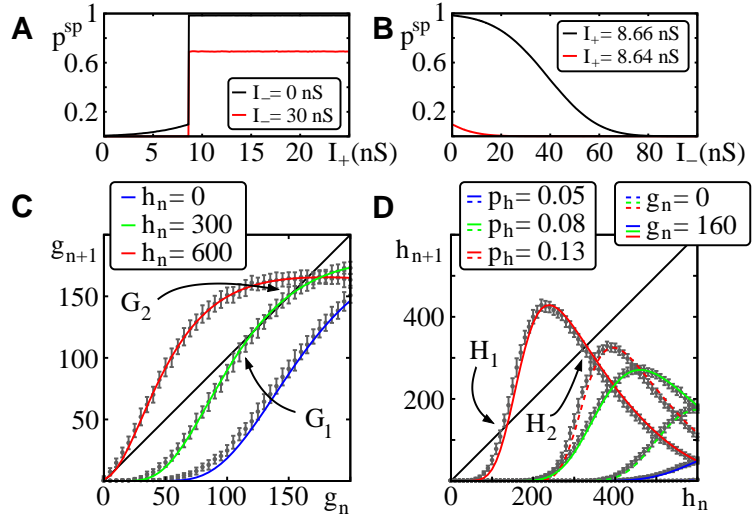}\caption{Hubs induce tangent bifurcations towards signal transmission (neuron
and network parameters as in Fig.~\ref{fig:Hubs-enable-propagation}).
(A,B): Firing probability $p^{\text{sp}}$ of a neuron in the ground
state as a function of synchronous (A) excitatory input $I_{+}$ and
(B) inhibitory input $I_{-}$. (C,D): Iterated maps for (C) the number
$g_{n}$ of (synchronously) active neurons in the $n$th subnetwork
(different colors indicate different fixed $h_{n}$) and (D) the number
of synchronized hub neurons $h_{n}$ (different colors: $p_{h}$ fixed;
different linestyles: $g_{n}$ fixed). Analytical predictions (solid/dashed
lines; Eqs.~(\ref{eq:gnp1},\ref{eq:hnp1})) agree well with numerical
simulations of the spiking neural network model (markers). Sufficiently large $h_{n}$ enables propagation
of synchrony (C) and a sufficiently large connection probability $p_{\text{h}}$
enables a persistent hub echo to a propagating synchronous pulse (D).
Hubs and non-hubs reactivate each other. \label{fig:threshold units}}
\end{figure}

To further understand this co-action mechanism, we consider the dynamics only
at the relevant time intervals where synchronous pulses are sent and
received. Observing that the neurons effectively act as  
probabilistic threshold units, we derive
an approximate analytic map for the joint response sizes of active
hubs and signal carrying (non-hub) units. The spiking probability due
to a synchronous input below the dendritic threshold $\Theta_{\text{d}}$
is very low (cf.~\ref{fig:threshold units}A), so that we neglect it against the probability of spiking
due to inputs above threshold. The probability $p^{\text{sp}}(I_{+},I_{-})$
of a neuron spiking in response to excitatory and inhibitory inputs
$I_{+}$ and $I_{-}$ is a function of the probability distribution
of the membrane potentials of that neuron at the time of input reception.
We take this dependency into account by assuming that immediately
before every spike reception time the neuronal state is distributed
as in the unperturbed ground state. 
The function $p^{\text{sp}}$ thus obeys
\begin{equation}
p^{\text{sp}}\left(I_{+},I_{-}\right)=\begin{cases}
0 & \text{for}\, I_{+}<\Theta_{\text{d}}\\
p^{0}\left(I_{-}\right) & \text{for}\, I_{+}\geq\Theta_{\text{d}}
\end{cases}\label{eq:psp}
\end{equation}
where $p^{0}\left(I_{-}\right)$ is the spiking probability of a neuron
in the ground state receiving a dendritically suprathreshold excitatory
input and an inhibitory input of size $I_{-}$. In particular, $p^{0}(0)$
is the spiking probability of a single neuron when a dendritic spike
is generated in the absence of inhibition. 
$p^0$ depends solely on the inhibitory input $I_{-}$, because on the one hand
only sufficiently strong excitatory inputs exceeding the dendritic threshold elicit a
dendritic spike and the effect of a dendritic spike on the postsynaptic neuron
saturates, i.e., it remains the same, for stronger excitation (cf.~Fig.~\ref{fig:threshold units}A),
as found in experiments \cite{DSExp}. 
On the other hand, 
inhibition will generally decrease a
neuron's spiking probability as it partially compensates the input
to the soma due to the dendritic spike (cf.~Fig.~\ref{fig:threshold units}B
and the experimental findings in \cite{Muller2012}).
The precise form of $p^{0}\left(I_{-}\right)$
depends on the details of the background activity and the properties
of neurons and interactions. 
As will become clear below, all \textit{qualitatively} similar $p^{0}\left(I_{-}\right)$
induce the same type of bifurcation relevant for robust signal transmission
and thus details of $p^{0}\left(I_{-}\right)$ do not matter. 

During robust signal transmission promoted by a hub echo, spikes of
hub neurons and neurons of the currently active subnetwork dominate
the network dynamics (cf.~inset of Fig.~\ref{fig:Hubs-enable-propagation}B).
We thus focus on these two groups of neurons. The influence of the
remaining neurons can be analytically derived analogously \cite{SI}.
To be specific, assume that $g_{n}\leq N_{\text{g}}$ neurons in a
given subnetwork $n$ and $h_{n}\leq N_{\text{h}}$ hub neurons are
active simultaneously, i.e., they spike synchronously. 
Given the random network topology, 
for sufficiently large $g_n$ and $h_n$ 
the total input to the neurons of the $(n+1)$th
subnetwork is approximately Gaussian distributed (approximating the actual Binomial distributions), $I_{+/-}\sim\mathcal{N}\left(\mu_{+/-},\sigma_{+/-}^{2}\right)$,
with probability density functions $f_{+}\left(I_{+}\right)$ and $f_{-}\left(I_{-}\right)$, and means and variances given by 
\begin{align}
\mu_{+} & \negthinspace=\negthinspace\negthinspace\left(\epsilon_{+}h_{n}+\epsilon_{\text{c}}g_{n}\right)p,\,\sigma_{+}^{2}\negthinspace=\negthinspace\left(\epsilon_{+}^{2}h_{n}+\epsilon_{\text{c}}^{2}g_{\text{n}}\right)p\left(1-p\right)\negthinspace,\label{eq:musigma+}\\
\mu_{-} & \negthinspace=\negthinspace\negthinspace\epsilon_{-}\left(h_{n}+g_{n}\right)p,\,\text{and}\,\sigma_{-}^{2}\negthinspace=\negthinspace\epsilon_{-}^{2}\left(h_{n}+g_{\text{n}}\right)p\left(1-p\right)\negthinspace.\label{eq:musigma-}
\end{align}
The expected number of neurons that spike synchronously in subnetwork
$n+1$ becomes 
\begin{equation}
g_{n+1}=N_{\text{g}}\int_{0}^{\infty}\negthinspace\negthinspace\negthinspace\int_{0}^{\infty}\negthinspace\negthinspace\negthinspace p^{\text{sp}}\left(I_{+},I_{-}\right)f_{+}\left(I_{+}\right)f_{-}\left(I_{-}\right)dI_{+}dI_{-}.\label{eq:iteratedmaplayer}
\end{equation}
 Whereas $p^{\text{sp}}$ discontinuously
depends on $I_{+}$, it changes smoothly and thus locally linearly
with $I_{-}$ (cf.~Fig.~\ref{fig:threshold units}A,B) such that
we may set $f_{-}\left(I_{-}\right)=\delta\left(I_{-}-\mu_{-}\right)$
to evaluate the integral in Eq.~(\ref{eq:iteratedmaplayer}), yielding
the iterated map
\begin{eqnarray}
g_{n+1} & = & N_{\text{g}}p^{0}\left(\mu_{-}\right)\frac{1}{2}\left(1+\text{Erf}\left[\frac{\Theta_{\text{d}}-\mu_{+}}{\sqrt{2}\sigma_{+}}\right]\right)\label{eq:gnp1}
\end{eqnarray}
for the number of active signal transferring (non-hub) neurons in
the next subnetwork. Note that all three quantities $\mu_{-}$, $\mu_{+}$
and $\sigma_{+}$ depend on $h_{n}$ and $g_{n}$ through Eqs. (\ref{eq:musigma+})
and (\ref{eq:musigma-}).

The iterated map for the number of synchronously active hub neurons
$h_{n+1}$ is derived analogously: We discard those $h_{n}$ neurons
that have spiked together with the $n$th subnetwork because they
are unlikely to  spike again due to their relative refractoriness,
such that $N_{\text{h}}-h_{n}$ hub neurons are available to spike.
Replacing $N_{\text{g}}$ by $N_{\text{h}}-h_{n}$ in Eq. (\ref{eq:iteratedmaplayer})
and computing the Gaussian probability densities of the inputs yields
the iterated map
\begin{equation}
h_{n+1}=\left(N_{\text{h}}-h_{n}\right)p^{0}\left(\tilde{\mu}_{-}\right)\frac{1}{2}\left(1\negthinspace+\negthinspace\text{Erf}\left[\frac{\Theta_{\text{d}}-\tilde{\mu}_{+}}{\sqrt{2}\tilde{\sigma}_{+}}\right]\right)\negthinspace,\label{eq:hnp1}
\end{equation}
where $\tilde{\mu}_{\text{+}}=\epsilon_{\text{+}}p_{\text{h}}\left(h_{n}+g_{n}\right)$,
$\tilde{\mu}_{-}=\epsilon_{-}p_{\text{h}}\left(h_{n}+g_{n}\right)$
and $\tilde{\sigma}_{\text{+}}^{2}=\epsilon_{\text{+}}^{2}p_{\text{h}}\left(1-p_{\text{h}}\right)\left(h_{n}+g_{n}\right)$. 

The joint two-dimensional map (\ref{eq:gnp1},\ref{eq:hnp1}) explicates
how the hub neurons can enable robust propagation of synchrony (see
Fig.~\ref{fig:threshold units}C,D): For a given number $h_{n}$ of
active hub neurons, the fixed points of Eq.~(\ref{eq:gnp1}) determine
whether robust propagation of synchrony can be initiated in the chain
of subnetworks. For networks without (active) hubs, $h_{n}=0$, there
is only one fixed point $G_{0}=0$ and any initial synchronous pulse
extinguishes after a small number of subnetworks. With increasing
$h_{n}$, two additional fixed points, $G_{1}$(unstable) and $G_{2}$(stable),
appear via a tangent bifurcation at some $h_{n}=h^{\ast}$ and robust
signal transmission is enabled for initial synchronous pulses $g_{0}\geq G_{1}$
(cf.~Fig.~\ref{fig:threshold units}C). For large numbers of active
hubs, even small initial group sizes $g_{0}$ are sufficient to generate
robust signal transmission across the chain of subnetworks.

Analogously, the fixed points of Eq.~(\ref{eq:hnp1}) determine whether
a persistent hub echo to the propagating synchronous pulse establishes
for a given hub connectivity $p_{\text{h}}$ and group size $g_{n}$
(cf.~Fig.~\ref{fig:threshold units}D). For small $p_{\text{h}}$
and $g_{n}$ there is only one fixed point $H_{0}=0$. With increasing
$p_{\text{h}}$ or $g_{n}$ two additional fixed points $H_{1}$ (unstable)
and $H_{2}$ (stable) appear via a tangent bifurcation for some $p_{\text{h}}^{\ast}$
and $g^{\ast}$. Thus, for sufficiently large hub connectivity
$p_{\text{h}}\geq p_{\text{h}}^{\ast}$, a persistent echo to a propagating
synchronous pulse of size $g_{n}$ can be established; equivalently,
for fixed connectivity $p_{\text{h}}$, sufficiently many synchronously
active neurons in the subnetwork maintain a hub echo. The bifurcations
resulting from the analytic mapping (\ref{eq:gnp1},\ref{eq:hnp1})
approximately predict the numerically found region where robust signal
transmission is possible (see dashed line in Fig.~\ref{fig:Hubs-enable-propagation}
and \cite{SI}). 

\begin{figure}[!tb]
\centering{}\includegraphics[width=0.6\columnwidth]{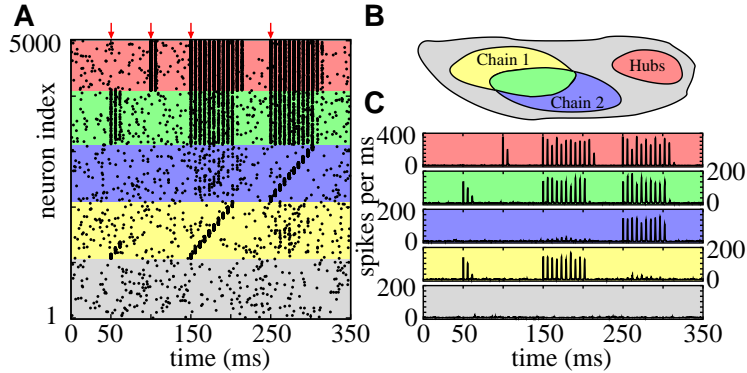}\caption{Hub-neurons
act as a generic signal amplifier and  activate
different signal routes. Figure shows simulation data for a sparse,
recurrent spiking neural network (same network as in Fig.~\ref{fig:Hubs-enable-propagation}B),
with two chains of subnetworks. (A) Rasterplot of the network activity;
the background colors indicate whether the neurons are members of
one or both chains, hub neurons or remaining neurons, as visualized
by (B). (C) Current activity (spikes per bin; bin size $1$ms) of the
different neuron populations. If synchronous spiking is initiated
either in the initial subnetwork of one chain ($t=50$ms) or the
hub neurons ($t=100$ms) only, synchronous activity extinguishes quickly.
In contrast, if the initial subnetwork of one of the chains as well
as the hub neurons are excited ($t=150$ms, $250$ms), robust propagation
of synchrony establishes in that specific chain. \label{fig:readout}}
\end{figure}

Having gained this mechanistic understanding, we now illustrate that
hubs unspecifically but selectively activate synchrony propagation.
Signal propagation becomes possible along any chain of subnetworks
that structurally exists in the system if its initial group is excited.
In particular, in systems with a second chain of subnetworks embedded,
the mutual hub/non-hub feedback can amplify signal transmission
along one chain without activating transmission in the other one (cf.~Fig.~\ref{fig:readout}).

In summary, we have demonstrated that hubs may act as amplifiers that
enable signal generation and transmission in recurrent networks.
So far, hubs were thought to themselves directly distribute various
types of signals (e.g., actual information in the world wide web,
certain infections in disease spreading, people in travel networks)
across a network. We now identified a complementary, fundamentally
different role of hubs in signal transmission: The hubs studied
here do not communicate the specific signal themselves; instead, increased
hub activity mirrors the presence of some localized signal in other
network parts and the hubs promote the transmission of any such signal
across sequences of non-hub subnetworks.

This mechanism of hub-activated signal propagation essentially relies on (a) the existence of some
highly connected nodes and (b) some sharp, threshold-like processing of incoming inputs by single
units (as for instance mediated by fast dendritic sodium spikes in neural circuits).
Furthermore, the phenomenon is robust against changes in the network topology. 
As explicit example we show that it occurs in scale-free networks networks \cite{Newman}, 
where hubs naturally emerge due to the ``fat-tail'' of the degree distribution (cf. \cite{SI} for an example). 
We thus expect that this type of signal transmission may well play a role in biological networks and 
even be exploited in self-organized solutions of technical communication networks \cite{Klinglmayr2012}. 

It has long been hypothesized that cortical neural networks
transmit signals via propagating synchronous spiking activity across
subnetworks connected in a feed-forward manner \cite{Synfire,Embed,DSSynchProp}.
The results above now suggest that hubs might enable robust propagation
of synchronous signals even in weak embedded feed-forward structures by echoing the synchronous signal propagating
along them. In the absence of hubs
(and due to the lack of a confining mechanism) the echo cannot contribute in this way as synchronous activity
either dies out or spreads across the whole network and causes pathological
activity (e.g., \cite{Embed} and cf.~also Fig. \ref{fig:Hubs-enable-propagation}A). 
To reveal the essential mechanisms underlying signal transmission, 
we disregarded ``Dale's Law'' \cite{Dale} (stating that each neuron either has only excitatory or only inhibitory outgoing connections) and 
considered a simple bimodal degree distributions clearly splitting the system into hub and non-hub neurons. 
In additional simulations, we verified that the uncovered new type of signal transmission equally emerges in networks with neurons 
obeying Dale's Law and exhibiting a natural and broad degree distributions \cite{SI}.

Interestingly, hubs have recently also been uncovered experimentally
in the developing hippocampus \cite{Bonifazi2009}. As in adult hippocampus,
synchronized oscillatory activity abounds and the structural feature
of hub neurons might support the directed transmission of specific
signals. Such hub-feedback support may provide one reason why hubs
emerge in these systems in the first place, cf. also \cite{Tattini2012}.

Specifically, hub-feedback might be also involved in the replay of
spike sequences during so-called sharp wave-ripple complexes observed
in the hippocampus \cite{Replay}.
Here, during sleep neurons are activated in the same order as they
have been during an exploration phase, accompanied by strong network
oscillations. Whereas most neurons take part in only a few of the
different replayed patterns, some are activated in a large fraction of events
\cite{Ylinen1995a}. Our results suggest that the latter may be 
  unspecific to certain memories and rather hub neurons
generating a synchronous feedback signal to stabilize signal propagation
along a previously learned feed-forward structure of specific neurons. 

Finally, our analytical results (\ref{eq:gnp1}) and (\ref{eq:hnp1})
for the activity of the hubs and the signal-carrying units clearly demonstrate
that the principle of mutual activation  underlying the support of signal transmission may
act in any network of sharply nonlinear (probabilistic) threshold
units, as characterizing, e.g., transmission of rate activities
in networks of neural populations (McCullogh-Pitts model, e.g., \cite{McP}), 
(failure) cascades in social, supply or communication networks (e.g., \cite{Cascades}), or
signaling in gene and protein networks (threshold Boolean networks, e.g.\, \cite{Bornholdt2008}).

This work was supported by the BMBF (grant no. 01GQ1005B) and the
DFG (grant no. TI 629/3-1). Simulation results were obtained using
the simulation software NEST \cite{Gewaltig2007}.

\bibliographystyle{plunsrt}

\end{document}